\documentclass[useAMS,usegraphicx,usenatbib]{mn2e}
\usepackage{graphicx}
\usepackage{amsmath}
\usepackage{amssymb}
\usepackage{subfigure}
\usepackage{url}
\usepackage{multirow}
\usepackage[linkcolor=blue,colorlinks=true,breaklinks,citecolor=blue,urlcolor=blue]{hyperref} 

\newcommand{\Mpc}{\rm\thinspace Mpc}

\newcommand{\km}{\rm\thinspace km}

\newcommand{\cm}{\rm\thinspace cm}

\newcommand{\s}{\rm\thinspace s}
\newcommand{\ks}{\rm\thinspace ks}

\newcommand{\Msun}{\hbox{$\rm\thinspace M_{\odot}$}}

\newcommand{\keV}{\rm\thinspace keV}

\newcommand{\erg}{\rm\thinspace erg}

\newcommand{\cts}{\rm\thinspace ct}
\newcommand{\ctsps}{\hbox{$\cts\s^{-1}\,$}}

\newcommand{\ergcmps}{\hbox{$\erg\cm\ps\,$}}

\newcommand{\kmps}{\hbox{$\km\s^{-1}\,$}}

\newcommand{\kmpspMpc}{\hbox{$\kmps\Mpc^{-1}\,$}}

\newcommand{\psqcm}{\hbox{$\cm^{-2}\,$}}
\newcommand{\pcmsq}{\hbox{$\cm^{-2}\,$}}

\newcommand{\ps}{\hbox{$\s^{-1}\,$}}

\newcommand{\rg}{\rm\thinspace $r_\mathrm{g}$}

\title[Flaring in Mrk~335]{Flaring from the supermassive black hole in Mrk~335 studied with \textit{Swift} and \textit{NuSTAR}}
\author[D. R. Wilkins et al.]{D. R. Wilkins$^1$\thanks{E-mail: drw@ap.smu.ca}\thanks{CITA National Fellow} L. C. Gallo$^1$, D. Grupe$^2$, K. Bonson$^1$, S. Komossa$^3$ and A.C. Fabian$^4$\\$^1$Department of Astronomy \& Physics, Saint Mary's University, Halifax, NS. B3H 3C3 Canada\\$^2$Space Science Center, Morehead State University, Morehead, KY 40351 USA\\$^3$Max-Planck-Institut f\"{u}r Radioastronomie, Auf dem H\"{u}gel 69, D-53121, Bonn, Germany\\$^4$Institute of Astronomy, University of Cambridge, Madingley Road, Cambridge. CB3 0HA UK}
\begin{document}

\date{Accepted 2015 September 11.  Received 2015 September 10; in original form 2014 April 8}

\pagerange{\pageref{firstpage}--\pageref{lastpage}} \pubyear{2015}

\maketitle

\label{firstpage}

\begin{abstract}
Monitoring of the narrow line Seyfert 1 galaxy Markarian 335 (Mrk~335) with the \textit{Swift} satellite discovered an X-ray flare beginning 2014 August 29. At the peak, the 0.5-5\keV\ count rate had increased from that in the low flux state by a factor of 10. A target of opportunity observation was triggered with \textit{NuSTAR}, catching the decline of the flare on 2014 September 20. We present a joint analysis of \textit{Swift} and \textit{NuSTAR} observations to understand the cause of this flare. The X-ray spectrum shows an increase in directly observed continuum flux and the softening of the continuum spectrum to a photon index of $2.49_{-0.07}^{+0.08}$ compared to the previous low flux observations. The X-ray spectrum remains well-described by the relativistically blurred reflection of the continuum from the accretion disc whose emissivity profile suggests that it is illuminated by a compact X-ray source, extending at most 5.2\rg\ over the disc. A very low reflection fraction of $0.41_{-0.15}^{+0.15}$ is measured, unexpected for such a compact corona. The X-ray flare is, hence, interpreted as arising from the vertical collimation and ejection of the X-ray emitting corona at a mildly relativistic velocity, causing the continuum emission to be beamed away from the disc. As the flare subsides, the base of this jet-like structure collapses into a compact X-ray source that provides the majority of the radiation that illuminates the disc while continuum emission is still detected from energetic particles further out, maintaining the low reflection fraction.

\end{abstract}


\section{Introduction}
Markarian 335 (Mrk~335) is a particularly fascinating example of a narrow line Seyfert 1 (NLS1) galaxy ($z=0.026$) whose X-ray emission has exhibited variability over more than an order of magnitude in flux over the last 15 years. Early observations from 2000 to 2006 found a bright X-ray source in a high flux state. Regular monitoring of this source with the \textit{Swift} satellite began in 2007 which found that the flux had dropped by a factor of 10 since the 2006 (and earlier) observations \citep{grupe+07}, hence a target of opportunity (ToO) observation in this low flux state was triggered with \textit{XMM-Newton} \citep{grupe+08}. Mrk~335 was found to again drop into a low flux state in 2009 but had recovered to an intermediate flux level by the time a new \textit{XMM-Newton} observation was triggered \citep{grupe+12}, with a further low flux state observed in 2013 which the source has remained in since. This extreme variability makes Mrk~335 an ideal target to study the physical processes through which energy is liberated from the accretion flow in active galactic nuclei (AGN) to power some of the most luminous objects in the Universe. The changes it undergoes giving rise to the extreme variability in its X-ray emission provide vital insight into the underlying processes.

X-ray spectra of Mrk~335, ranging from the early high-flux observations with \textit{XMM-Newton} \citep{crummy+06}, \textit{Suzaku} \citep{larsson+08} and even \textit{ASCA} \citep{ballantyne+01} to the low \citep{grupe+08,parker_mrk335,gallo+14} and intermediate flux \citep{gallo+13} observations  show X-ray continuum emission from a corona of energetic particles surrounding the central black hole in Mrk~335 illuminating the accretion disc of material spiralling inward \citep{george_fabian}. This leads to X-ray reflection by the processes of Compton scattering, photoelectric absorption and the emission of fluorescence lines and bremsstrahlung \citep{ross_fabian}. The reflection spectrum, including the prominent K$\alpha$ emission line of iron at 6.4\keV, is blurred by Doppler shifts and relativistic beaming due to the orbital motion of material in the accretion disc as well as by gravitational redshifts in the strong gravitational field around the black hole \citep{fabian+89,laor-91}.

Further compelling evidence for relativistically blurred reflection from the accretion disc in Mrk~335 comes from the detection of X-ray reverberation time lags between the variability in the X-ray continuum and the corresponding variations in the reflected X-rays \citep{kara+13}. The measured time lag corresponds to the light crossing time between the X-ray emitting corona and the inner regions of the accretion disc, while \citet{gallo+14} show that variation in the X-ray continuum illuminating the disc naturally explains the observed spectral changes between the high and low flux states.

In addition to the blurred reflection from the accretion disc, absorption features are seen in the X-ray spectrum that are attributed to variable outflows of material from the central regions of the AGN. \citet{longinotti+13} find significant spectral features due to absorption by ionised outflowing material, attributed to a wind launched from the surface of the accretion disc during the 2009 intermediate flux epoch and that some of this absorption remains during the 2006 and 2013 high and low flux epochs observed by \textit{XMM-Newton}.

Through analysis of the relativistically blurred reflection from the accretion disc, \citet{mrk335_corona_paper} found that the transition from the high to low flux states was accompanied by a contraction of the X-ray emitting corona from an extended structure to a confined region close to the black hole. One of the high flux epochs observed by \textit{Suzaku} in 2006 corresponded to a vertical collimation of the corona into a jet-like structure perpendicular to the plane of the accretion disc. They find that a flare during the 2013 low flux epoch during which the the X-ray count rate doubled for $\sim 100$\ks\ corresponded to a reconfiguration of the corona from a slightly extended to a much more compact structure with the flare itself caused by a brief vertical collimation of the corona, reminiscent of an aborted jet launching event.

\textit{Swift} monitoring of Mrk~335 revealed an X-ray flare in 2014 September lasting approximately 30 days \citep{komossa+15}. At the peak of the flare lasting around 10 days, the X-ray count rate increased by an order of magnitude upon which a ToO observation was triggered with the \textit{NuSTAR} hard X-ray imaging telescope. We here present analysis of the \textit{Swift} and \textit{NuSTAR} observations to understand the cause of the flare.

\section{Observations and Data Reduction}

\begin{figure}
\centering
\includegraphics[width=85mm]{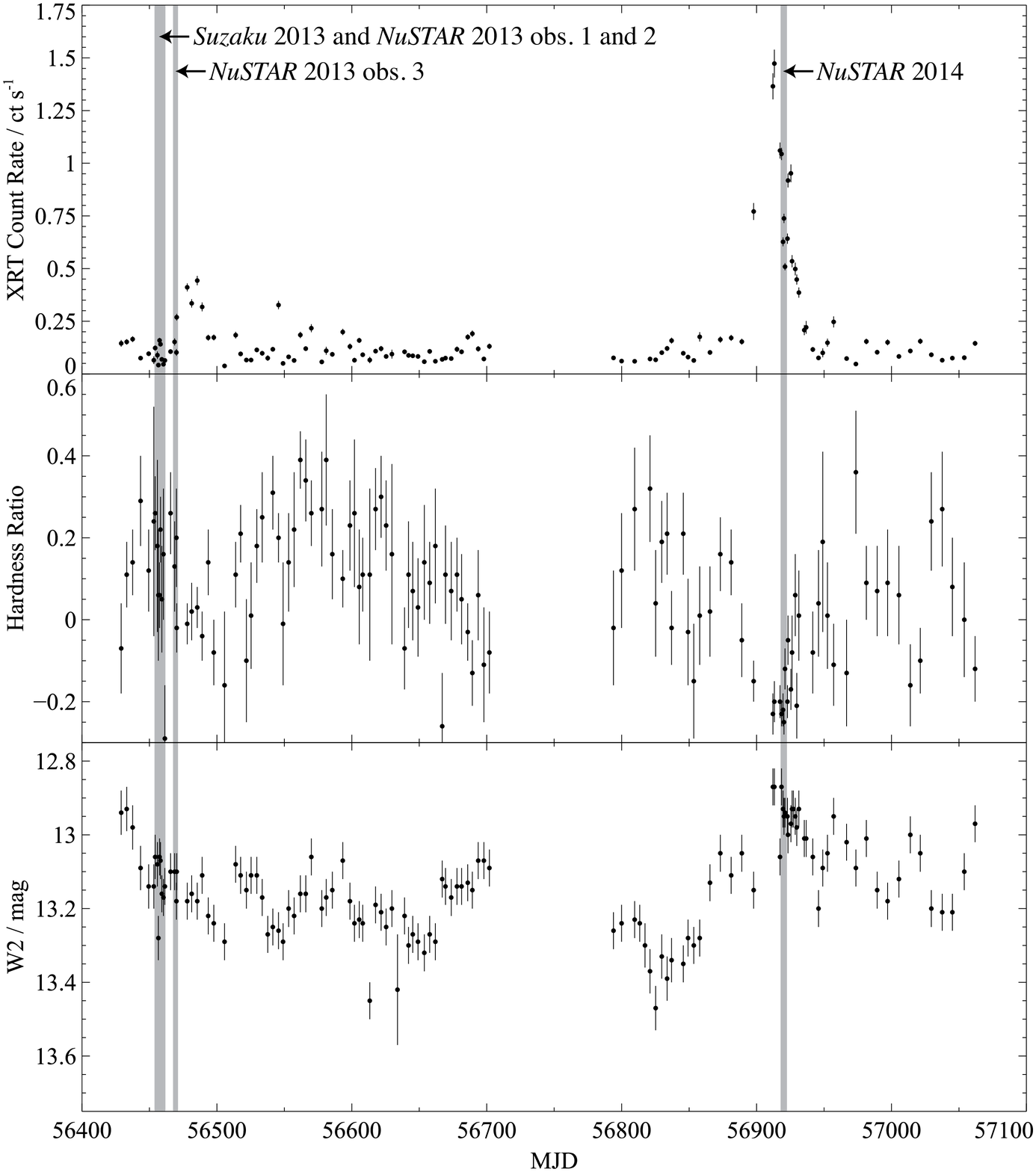}
\caption[]{The light curve of Mrk~335 recorded by \textit{Swift} between 2013 May 17 and 2015 February 09, showing the count rate measured in the \textit{X-ray telescope (XRT)}, the X-ray hardness ratio defined as $(H-S)/(H+S)$, where $H$ and $S$ are the count rates measured in the 1-10\keV\ and 0.3-1\keV\ energy bands respectively, and the ultraviolet flux measured through the \textit{UVW2} filter ($\lambda_0=2030$\,\AA) in the \textit{Ultraviolet and Optical Telescope (UVOT)}. Light curves are binned over each pointing made with \textit{Swift}. Indicated are the timings of the 2013 observations made with \textit{Suzaku} \citep{gallo+14,mrk335_corona_paper} and \textit{NuSTAR} \citep[three observations,][]{parker_mrk335} and the timing of the \textit{NuSTAR} ToO observation on the decline of the observed flare.}
\label{swift_lightcurve.fig}
\end{figure}

Mrk~335 is subject to regular monitoring with \textit{Swift}, with a $\sim 1$\ks\ observation taken weekly using the \textit{X-ray Telescope (XRT)} and \textit{Ultraviolet and Optical Telescope (UVOT)}. The light curve of Mrk~335 between 2013 May 17 and 2015 February 09 recorded with both \textit{Swift XRT} and \textit{Swift UVOT} using the \textit{UVW2} filter with central wavelength 2030\,\AA\ (and full width at half maximum, FWHM, 760\,\AA) is shown in Fig.~\ref{swift_lightcurve.fig}. Previous light curves of this source are shown in \citet{grupe+08} and \citet{grupe+12}.

A sharp increase in X-ray flux was noted on 2014 September 12 (MJD 56912) with the X-ray count rate increasing by more than an order of magnitude. The frequency of \textit{Swift} monitoring was increased with $\sim 1$\ks\ observations taken daily and a 75\ks\  ToO observation was triggered with \textit{NuSTAR} \citep{nustar} on 2014 September 20 (\textit{NuSTAR} OBSID 80001020002), although by this time the X-ray flare had begun to subside with the X-ray count rate having decreased to a third of its maximum. This is the brightest Mrk~335 has been seen since \textit{Swift} monitoring began in 2007.

\begin{figure}
\centering
\includegraphics[width=85mm]{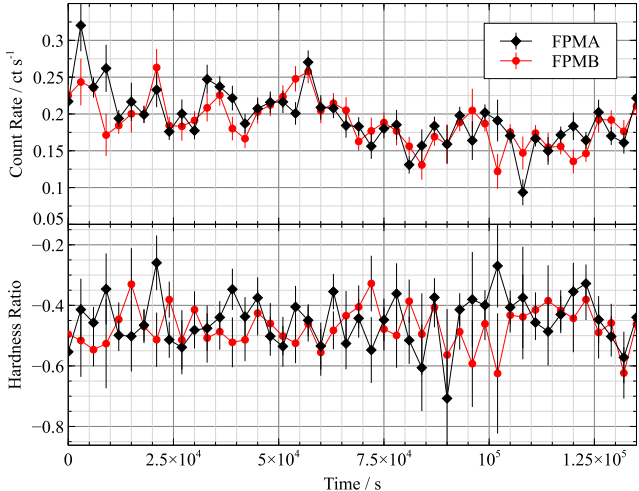}
\caption[]{The background-subtracted light curve of Mrk~335 during the 75\ks\ \textit{NuSTAR} observation on the decline of the flare in 2014 showing the count rates from the source in both the FPMA and FPMB detectors (3-50\keV) in 3\ks\ time bins as well as the hardness ratio, $(H-S)/(H+S)$, between the 10-50 and 3-10\keV\ energy bands.}
\label{nustar_lc.fig}
\end{figure}

\subsection{NuSTAR}

The total effective exposure of the \textit{NuSTAR} observation was 69\ks\ with an average count rate of 0.17\ctsps\ detected in the Focal Plane Module A detector (FPMA) and 0.16\ctsps\ in the FPMB detector. The light curve of Mrk~335 during the observation on the decline of the flare is shown in Fig.~\ref{nustar_lc.fig} and reveals only a subtle decrease in the 3-50\keV\ count rate (less than 25 per cent) on the time-scale of this observation. The hardness ratio remains constant over the course of the observation suggesting that there is no significant variability in the spectral shape during this period.

Event lists from \textit{NuSTAR} were screened and reprocessed using the latest calibration following the standard procedure with \textsc{nustardas} v1.4.1. The source spectra were extracted from regions 60\,arcsec in diameter from both detectors, FPMA and FPMB, centred on the point source and corresponding background spectra were extracted from regions of the same size from regions of the detectors off away from the point source. Response matrices (the RMF photon redistribution matrix, or response matrix file and ARF ancillary response file describing the instrument effective area as a function of X-ray energy) were produced with the \textsc{nuproducts} pipeline used to extract the spectra. The spectra were binned using the \textsc{grppha} tool such that there were at least 20 counts in each spectral bin and that the errors are approximately Gaussian and that the $\chi^2$ statistic can be used to assess the goodness of fit of spectral models.

\subsection{Swift}
In order to extend spectral coverage below the 3\keV\ lower limit of \textit{NuSTAR}, the X-ray spectrum was extracted from the 1.4\ks\ \textit{Swift XRT} observation simultaneous with an early part of the \textit{NuSTAR} observation (\textit{Swift} OBSID 00033420004). The event list was reprocessed using the latest calibration and filtering criteria applied following the standard procedure using \textsc{xrtpipeline}.

\textit{Swift XRT} was operated in Windowed Timing mode for the observations in question, in which the CCD detector is continuously read out along one row, thus the image appears as a one dimensional line across the detector in which the photons accumulated from each column are summed into a single pixel. The source spectrum was extracted from a rectangular region spanning $40\times30$ pixels, aligned along the line of the image and centred on the bright point source. The corresponding background spectrum was extracted from a region the same shape and size from a different position along the line. The ancillary response matrix (ARF) was generated for the observation using \textsc{xrtmkarf} and the photon redistribution matrix (RMF) appropriate for the time and configuration of the observation was obtained from the \textsc{caldb}. The spectra were binned using the \textsc{grppha} tool such that there were at least 20 counts in each bin.

In order to determine the variability of the X-ray emission over the course of the longer \textit{NuSTAR} observation, \textit{Swift} snapshots were also analysed from the day before (0.6\ks, 2014 September 19, OBSID 00033420003) and after (1.6\ks, 2014 September 21, OBSID 00033420005). These observations were also conducted with the \textit{XRT} operating in Windowed Timing mode and the data were reduced following the same procedure.

The ultraviolet light curve was produced using the \textit{UVOT} through the \textit{UVW2} filter ($\lambda_0=2030$\,\AA) following the procedure of \citet{grupe+07}.

\section{Spectral Analysis}

\subsection{NuSTAR}
\begin{figure*}
\begin{minipage}{175mm}
\centering
\subfigure[] {
\includegraphics[width=80mm]{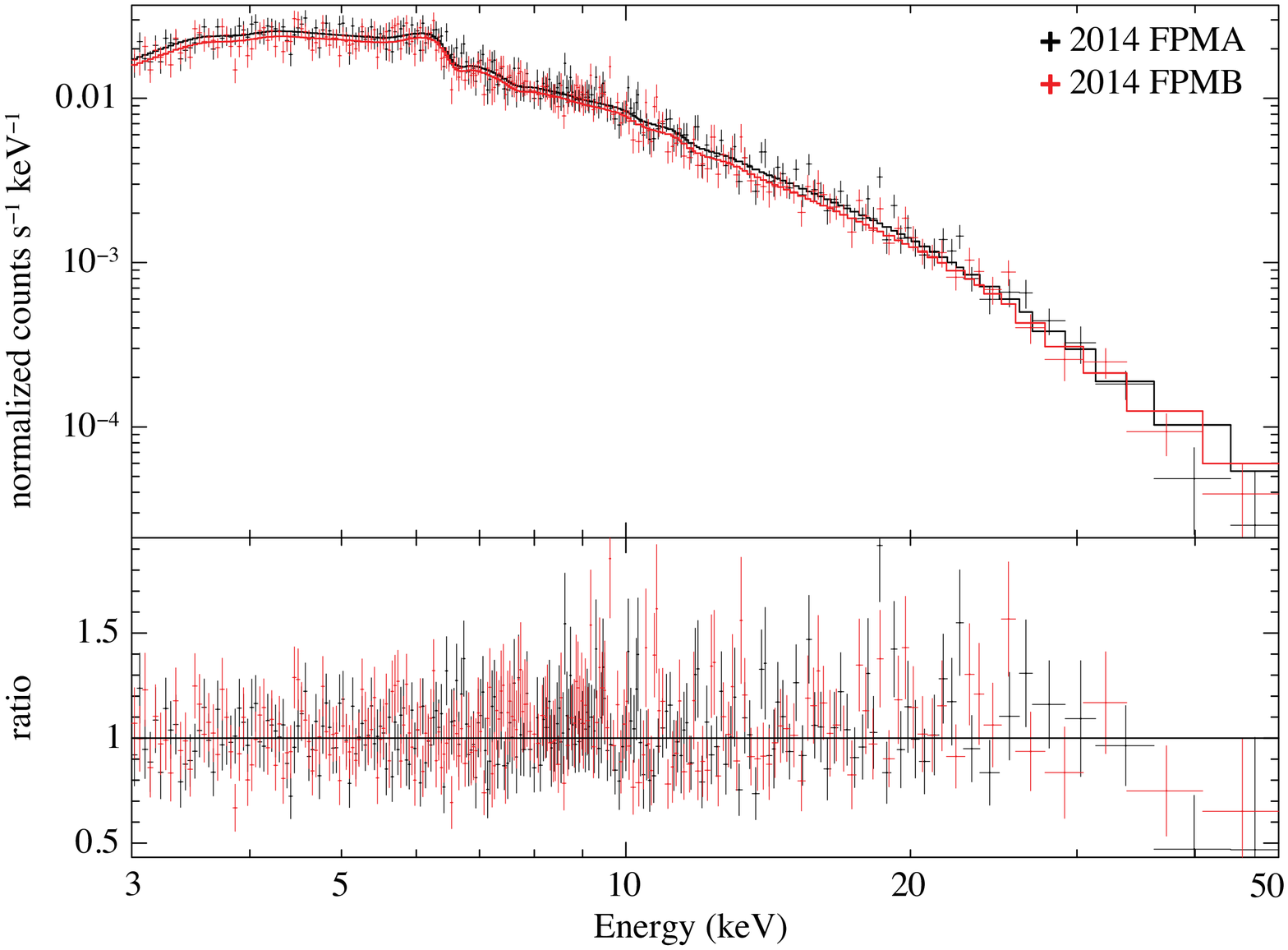}
\label{spectra.fig:nustar_2014}
}
\subfigure[] {
\includegraphics[width=80mm]{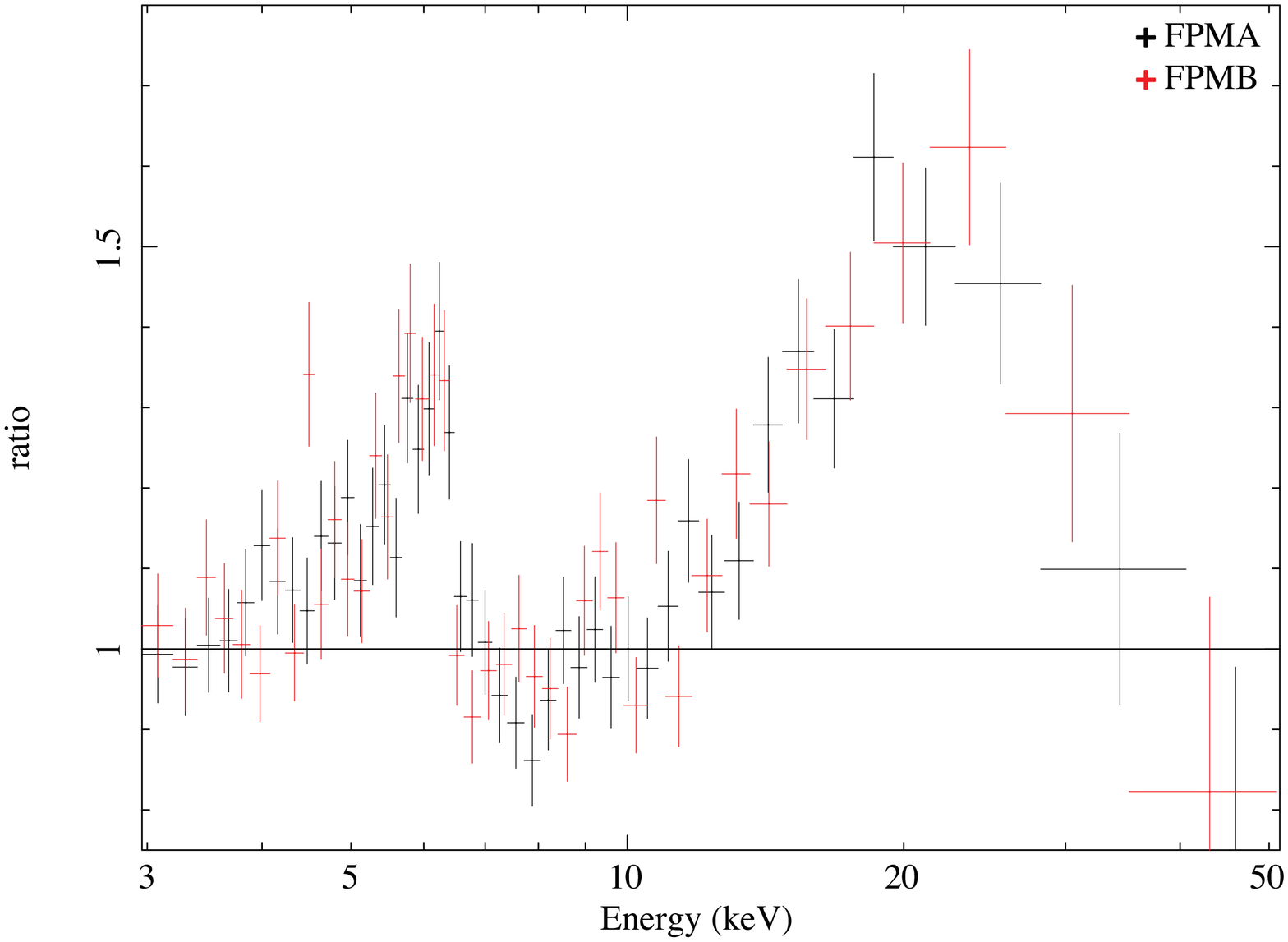}
\label{spectra.fig:nustar_plfit}
}
\subfigure[] {
\includegraphics[width=80mm]{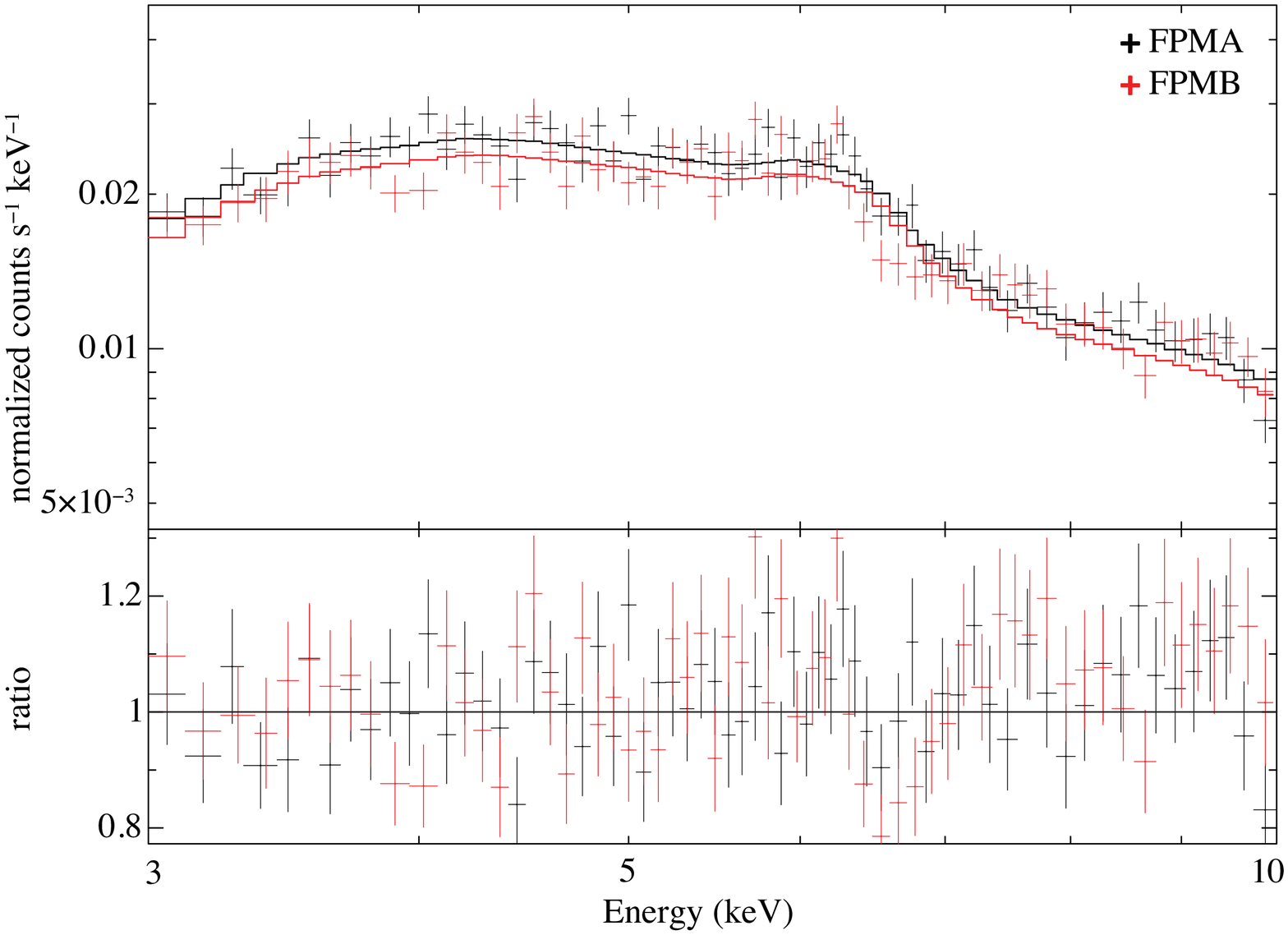}
\label{spectra.fig:fekresiduals}
}
\subfigure[] {
\includegraphics[width=80mm]{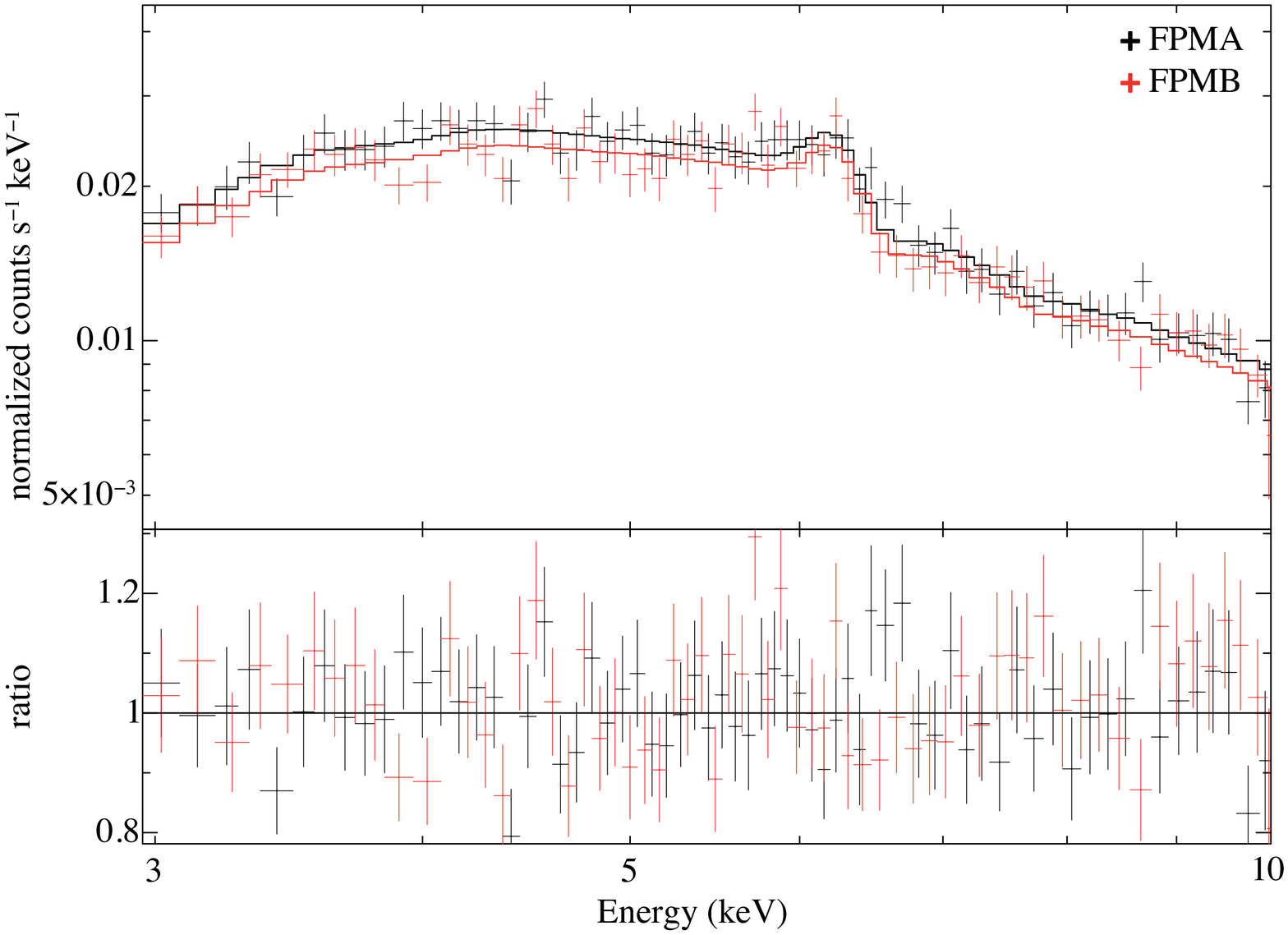}
\label{spectra.fig:fek}
}
\caption[]{\subref{spectra.fig:nustar_2014} The X-ray spectra of Mrk~335 measured by the \textit{NuSTAR} FPMA and FPMB detectors during the decline of the 2014 flare with the best-fitting model comprising the relativistically blurred reflection of a power law X-ray continuum from the accretion disc. \subref{spectra.fig:nustar_plfit} The ratio of the 2014 \textit{NuSTAR} spectrum to the best-fitting power law continuum (fit over the energy range 8-10\keV), showing the relativistically broadened iron K$\alpha$ emission line at 6.4\keV\ and `hump' due to Compton scattering from the accretion disc around 20\keV. \subref{spectra.fig:fekresiduals} The 2014 \textit{NuSTAR} spectra over the 3-10\keV\ energy band fit with just power law continuum emission and relativistically blurred reflection from the accreiton disc, showing the residuals due to the narrow core of the iron K$\alpha$ line at 6.4\keV\ and absorption around 7\keV. \subref{spectra.fig:fek} The 2014 \textit{NuSTAR} spectra fit over the 3-10\keV\ energy band including unblurred reflection from distant material and absorption from the outflowing disc wind.}
\label{spectra.fig}
\end{minipage}
\end{figure*}

\begin{figure}
\centering
\includegraphics[width=85mm]{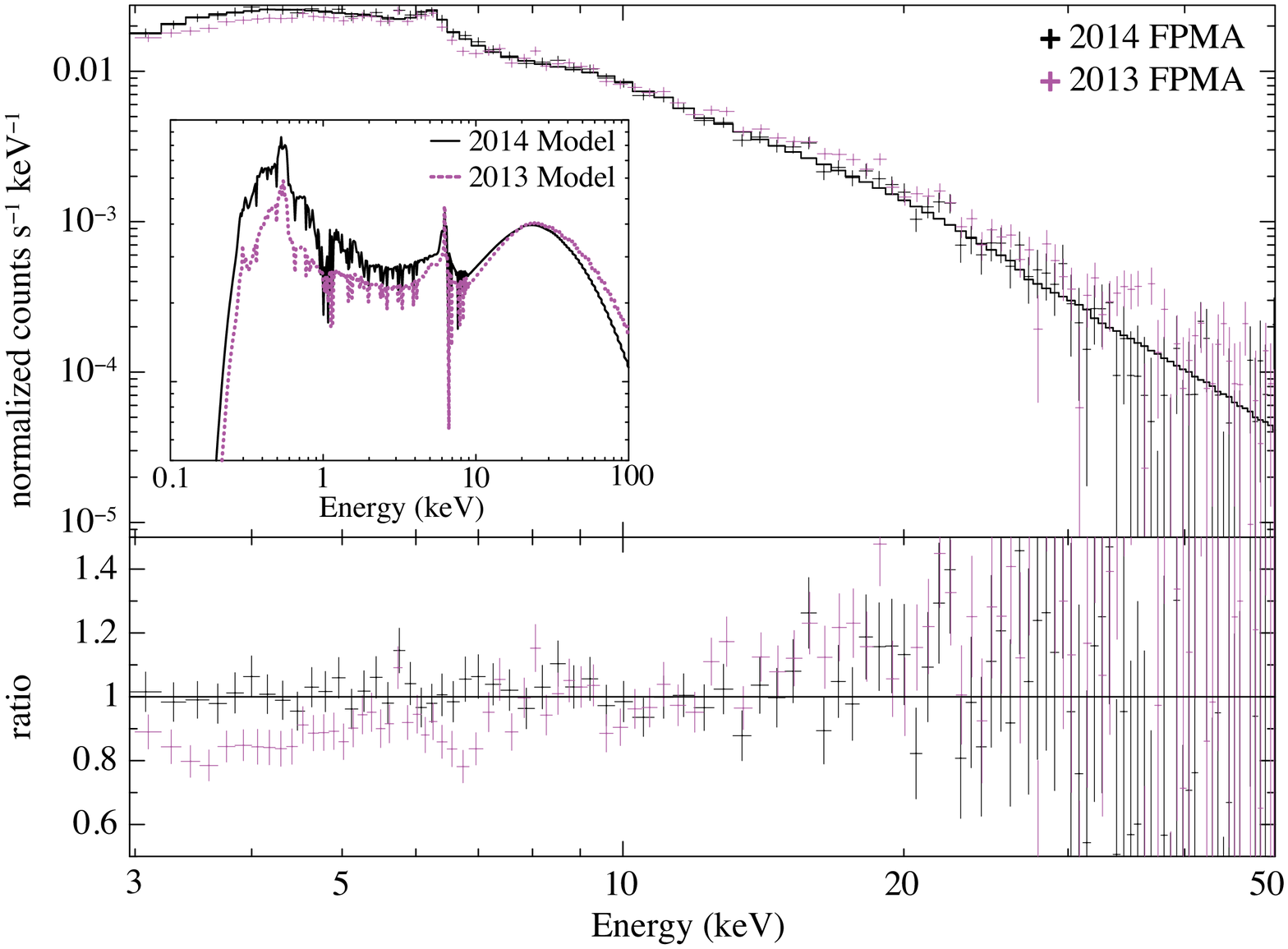}
\caption[]{The 2014 \textit{NuSTAR} spectra compared to the third observation made in 2013 along with the best fitting model to the 2014 spectra. The inset compares the best-fitting models to the 2013 and 2014 spectra, showing the change in spectral slope between the two observations.}
\label{nustar_2013_2014.fig}
\end{figure}

The \textit{NuSTAR} spectrum during the subsidence of the flare is shown in Fig.~\ref{spectra.fig:nustar_2014} and is remarkably similar to that at the highest flux level seen during the low state by \textit{NuSTAR} in 2013 \citep{parker_mrk335} except with a slight excess in emission seen below 5\keV\ in the 2014 observation, apparent in Fig.~\ref{nustar_2013_2014.fig}. The X-ray spectra were analysed using \textsc{xspec} \citep{xspec}.

We are motivated by the finding that the spectrum during the previous low flux state of Mrk~335 is well-described by the relativistically blurred reflection of the X-ray continuum from the accretion disc \citep{parker_mrk335,gallo+14} and that the variability seen in the X-ray spectrum between the low and high flux epochs is naturally explained by the reflection of a variable X-ray continuum from an expanding corona \citep{gallo+14,mrk335_corona_paper} to model the 2014 spectrum in terms of this blurred reflection and to understand the cause of the flare in such a scenario. Indeed, the relativistically broadened iron K$\alpha$ emission line at 6.4\keV\ and the `hump' in the reflection spectrum due to Compton scattering around 20\keV\ remain apparent \citep[as in][]{parker_mrk335} in the ratio between the \textit{NuSTAR} spectrum and best-fitting power law shown in Fig.~\ref{spectra.fig:nustar_plfit}. Indeed, including relativistically blurred reflection from the accretion disc in the spectral model improves the fit to the \textit{NuSTAR} spectrum over just a simple power law with $\Delta\chi^2 = 221$ (for 621 degrees of freedom). The parameters of the blurred reflection model are kept free such that while maintaining the scenario suggested by previous observations, the model is not bound by the previous configuration of the illuminating corona and accretion disc.

The \textit{NuSTAR} spectrum was initially fit with a simple model consisting of just continuum emission directly observed from the corona, represented by a power law, plus the relativistically blurred reflection of such a continuum spectrum from the accretion disc. The relativistically broadened reflection spectrum is represented by the \textsc{reflionx} model of \citet{ross_fabian}  including reflection through photoelectric absorption, fluorescent line emission, Compton scattering and thermal bremsstrahlung convolved with the relativistically broadened emission line model of \citet{laor-91} using \textsc{kdblur2} to fit the accretion disc emissivity profile as a once-broken power law. The continuum photon index and the inner radius, iron abundance and ionisation parameter ($\xi = 4\pi F/n$ for material with hydrogen number density $n$ irradiated by ionising flux $F$) were fit as free parameters. We find that the lower bound of the accretion disc inclination to the line of sight is not well constrained by these data, hence this is frozen to the value found by \citet{mrk335_corona_paper}, simultaneously fitting \textit{XMM-Newton} and \textit{Suzaku} observations of Mrk~335 between 2006 and 2013, since we do not expect this to have changed. The photon index of the continuum irradiating the disc to produce the reflection spectrum is tied to that of the directly observed continuum. Untying the direct and irradiating photon indices, we find that that the data do not require there to be any variation between these two parameters. There is, however, some degree of degeneracy between the spectral slope of the primary continuum and the reflection spectrum hence both photon indices are less well constrained when untied.

We find that the \textit{NuSTAR} spectrum between 3 and 50\keV\ is well described by this simple model ($\chi^2 / \nu = 1.02$ with 618 degrees of freedom), however focusing on just the 3-10\keV\ energy band, shown in Fig.~\ref{spectra.fig:fekresiduals}, encompassing the relativistically broadened iron K$\alpha$ fluorescence line at 6.4\keV\ and the associated photoelectric absorption edge at 7.1\keV, we find that it is necessary to include unblurred reflection from distant material to properly account for the narrow core of the iron K$\alpha$ emission line seen on top of the broad line. This is also modelled using \textsc{reflionx} with the iron abundance frozen to the solar value and the ionisation parameter frozen to 1\ergcmps, allowing only the normalisation to vary freely. Including the unblurred reflection improves the fit to the \textit{NuSTAR} spectrum giving $\Delta\chi^2 = 10$ (with one additional free parameter). In addition, it is necessary to include absorption from outflowing material consistent with the low flux state observed by \textit{Suzaku} in 2013 \citep{gallo+14} to account for the residuals around 7\keV. This is the highest ionisation state of the three outflowing absorption components found during the 2009 intermediate flux observation of Mrk~335 and is modelled using the photoionisation code \textsc{xstar} \citep{longinotti+13}. In addition to the existing free parameters, the column density and ionisation of the outflow are allowed to vary while the velocity is fixed at 5000\kmps\ to reproduce the observed energy of the absorption feature. Inclusion of absorption by the outflow further improves the fit with $\Delta\chi^2 = 16$ (with an additional 2 free parameters). The final fit to this band is shown in Fig.~\ref{spectra.fig:fek}.

Comparing the 2014 \textit{NuSTAR} observation to that taken in 2013 (Fig.~\ref{nustar_2013_2014.fig}), we see that although the spectra are largely consistent, there was a slight pivoting of the spectrum around 10\keV\ with the 2013 count rate slightly lower below this energy and slightly higher above when compared to the best-fitting model to the 2014 spectrum (but only by about $2\sigma$). This subtle change to the \textit{NuSTAR} spectrum is entirely explained by the softening of the continuum spectrum from $\Gamma\sim 2.3$ in 2013 \citep{parker_mrk335} to $\Gamma=2.43$ in 2014. The residual at 7\keV\ in the 2013 spectrum against the 2014 model suggests a change in the intrinsic absorption. We find that the best-fitting column density of the warm absorber was greater in 2013 at $(2.4_{-1.2}^{+1.7})\times 10^{23}$\psqcm\ compared to $(1.4_{-0.4}^{+1.4})\times 10^{23}$\psqcm\, though they are consistent within errors at the 90 per cent confidence level, thus the change is not statistically significant. This slight change to the intrinsic absorption is insufficient to alter the shape of the continuum seen in the \textit{NuSTAR} bandpass.

The ionisation parameter of the reflecting material in the accretion disc is poorly constrained by the \textit{NuSTAR} observation alone since the detector bandpass does not include the soft excess X-ray emission consisting of a number of emission lines below 1\keV. The flux from these lines increases greatly as the disc becomes more ionised. The reflection fraction is defined as the ratio of the reflected to directly observed continuum flux, extrapolated over the broad 0.1-100\keV\ energy band. We note that this is contrary to the definition used by \citet{parker_mrk335} who measure the fluxes over the 20-40\keV\ energy band, though the 0.1-100\keV\ reflection fraction allows the measurement to constrain the extent of the X-ray emitting corona discussed by \citet{mrk335_corona_paper}. We also compute reflection fractions over the energy range 20-40\keV\ for comparison with other work. The measured value of the 0.1-100\keV\ reflection fraction depends on the ionisation parameter of the disc which influences the total reflected flux inferred below 1\keV. Fitting the \textit{NuSTAR} data alone with no measurement of the soft excess is only able to constrain the ionisation parameter to $\xi < 53$\ergcmps\ through the non-detection of the K$\alpha$ emission line from highly ionised helium-like and hydrogenic iron at 6.67 and 6.97\keV\ respectively, which admits 0.1-100\keV\ reflection fractions between 0.13 and 1.34.

\subsection{Swift and NuSTAR simultaneous fit}

In order to constrain the ionisation parameter and hence the reflection fraction over the range 0.1-100\keV, we fit the 3-50\keV\ 75\ks\ \textit{NuSTAR} spectrum simultaneously with the 1.4\ks\ spectrum over the energy range 0.3-5\keV\ recorded by \textit{Swift XRT}, concurrent with the first part of the \textit{NuSTAR} observation. What is required is to constrain the upper limit to the flux over the energy range 0.5-1\keV, since the X-ray emission in fluorescence lines over this range increases as the disc becomes more ionised. An upper limit to the 0.5-1\keV\ X-ray flux will hence constrain the required upper limits to the ionisation parameter and reflection fraction.

\begin{figure}
\subfigure[] {
\includegraphics[width=80mm]{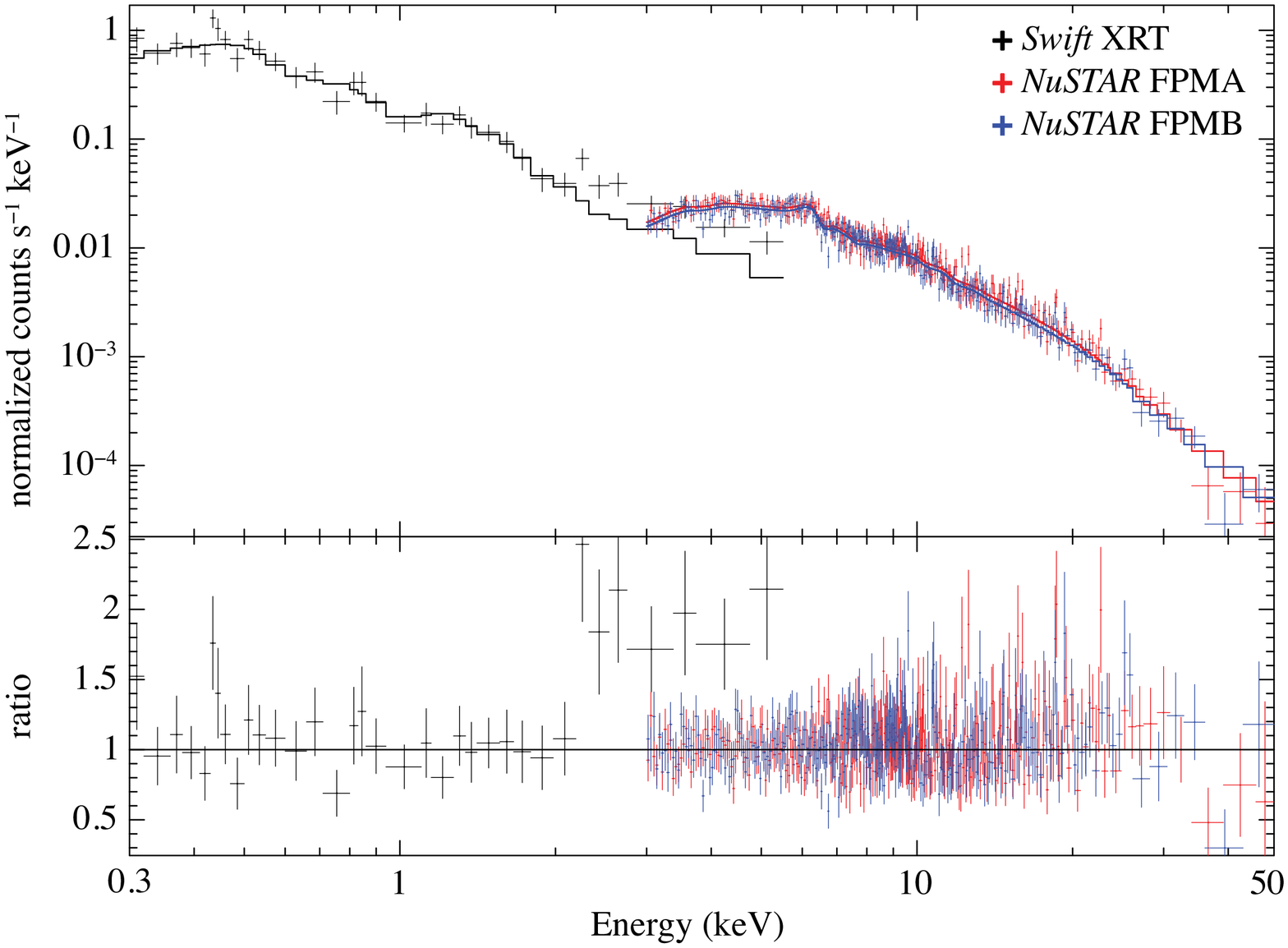}
\label{simfit.fig:const}
}
\subfigure[] {
\includegraphics[width=80mm]{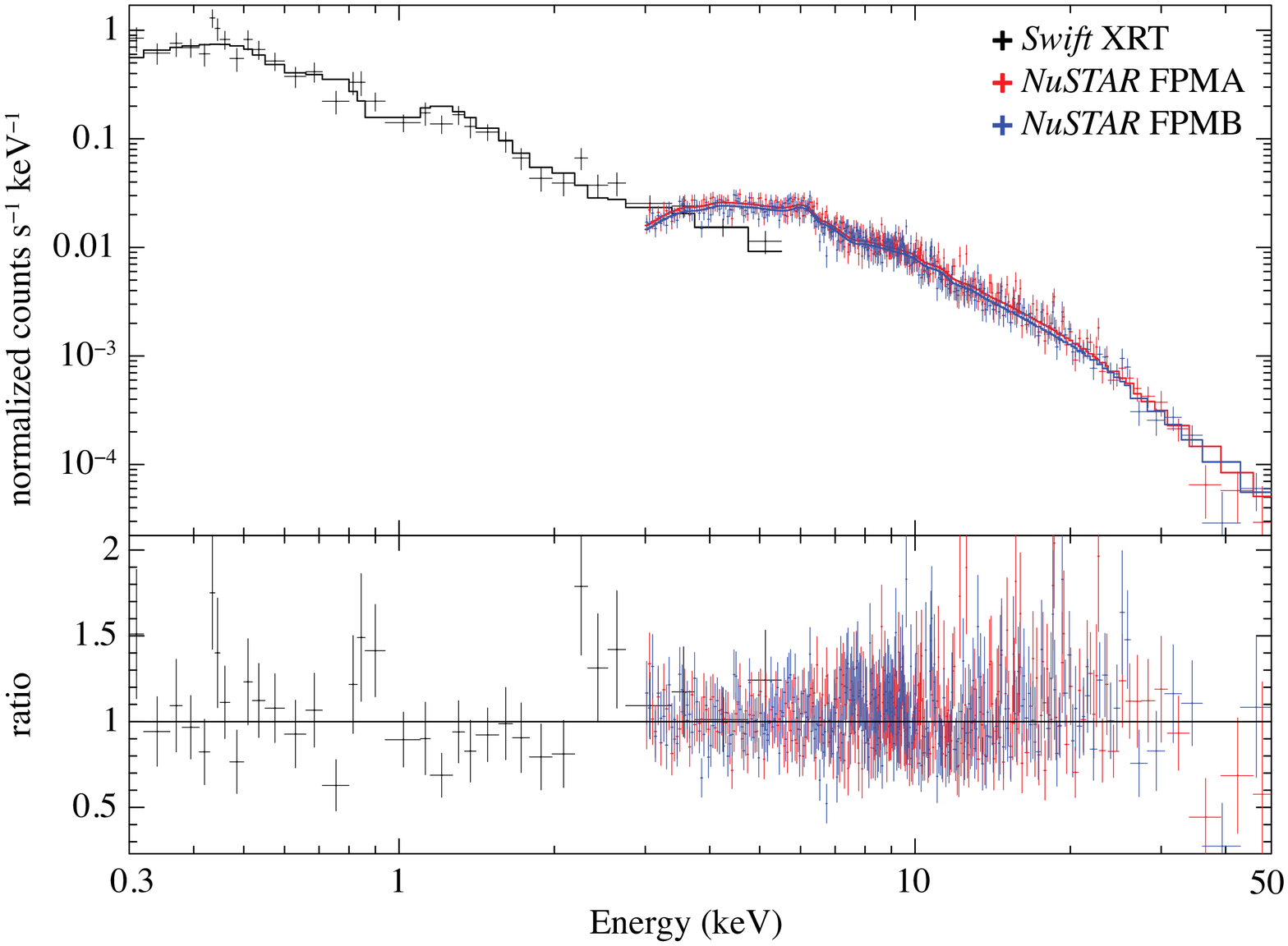}
\label{simfit.fig:noconst}
}
\subfigure[] {
\includegraphics[width=80mm]{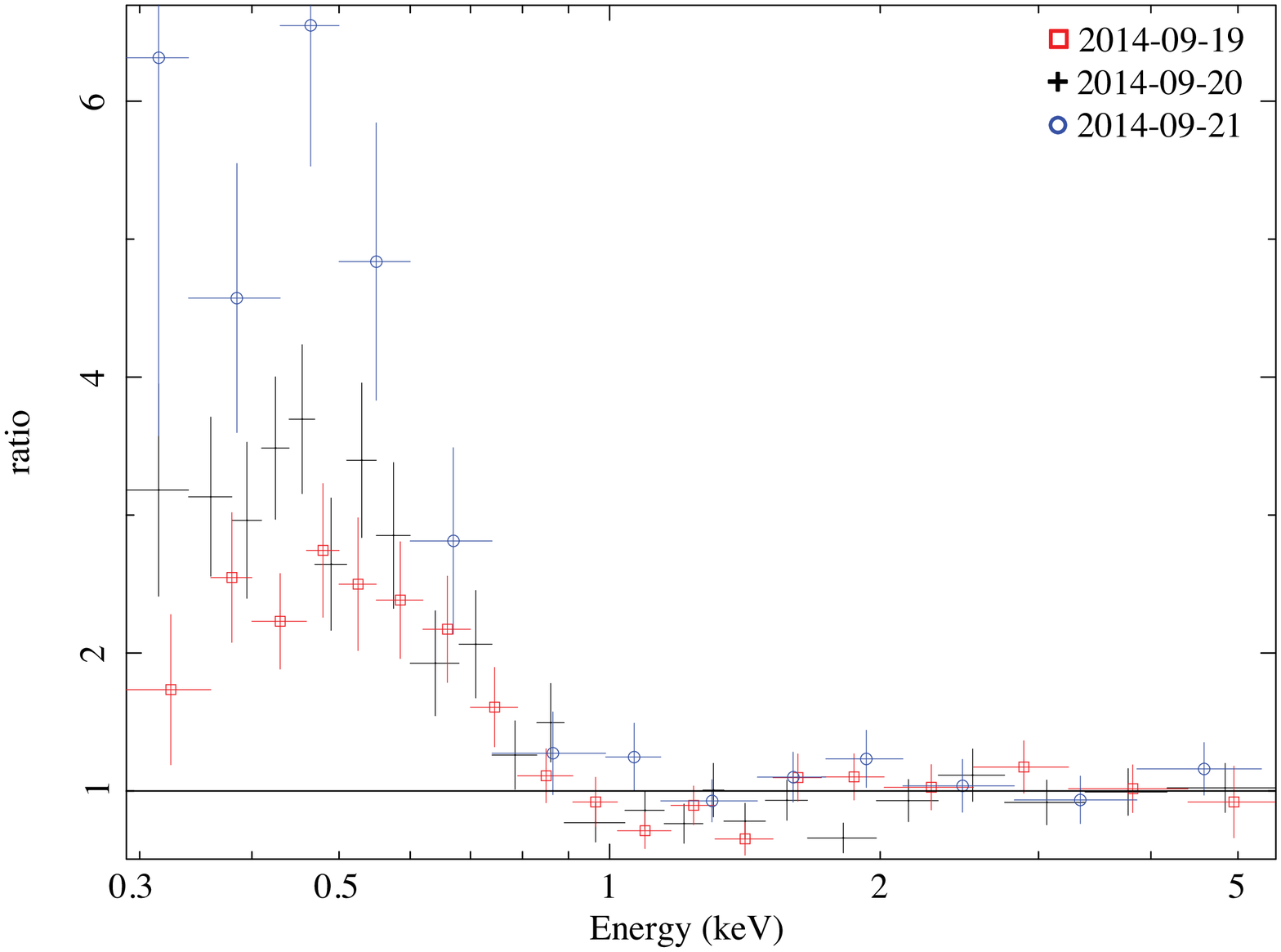}
\label{simfit.fig:ratios}
}

\caption[]{The 75\ks\ 2014 \textit{NuSTAR} spectrum fit simultaneously with the coincident 1.4\ks\ \textit{Swift XRT} spectrum with a model comprising the relativistically blurred reflection of a power law X-ray continuum from the accretion disc \subref{simfit.fig:const} initially fitting a cross normalisation constant between the \textit{Swift} and \textit{NuSTAR} spectra and \subref{simfit.fig:noconst} freezing the cross normalisation constant to unity to understand the origin of the discrepancy between 2.5 and 5\keV\ in terms of the evolution of the soft excess reflected from the accretion disc over the course of the longer \textit{NuSTAR} observation. \subref{simfit.fig:ratios} The ratios of the \textit{Swift XRT} spectra taken on the day of the flare (2014 September 20), the day before and the day after to the best-fitting power law to the \textit{Swift} spectrum above 2\keV\ on each day showing the evolution of the soft excess on the decline of the flare.}
\label{simfit.fig}
\end{figure}

We initially fit the same model with a freely varying multiplicative constant between the \textit{NuSTAR} and \textit{Swift} to allow for a systematic offset in the absolute flux calibration between the two instruments. The result is shown in Fig.~\ref{simfit.fig:const}. We find that the model describes both the \textit{Swift XRT} and \textit{NuSTAR} spectra well ($\chi^2 / \nu = 1.02$ with 668 degrees of freedom). Through the cross normalisation constant, the count rate in the \textit{NuSTAR} spectrum exceeds that predicted by the \textit{Swift} model by a factor of $1.3_{-0.2}^{+0.1}$, compared to a factor close to unity found in previous simultaneous studies between \textit{Swift} and \textit{NuSTAR} \citep[\textit{e.g.}][]{parker_mrk335}. There is, however, a discrepancy between the \textit{Swift} and \textit{NuSTAR} data (and the best-fitting model) with the \textit{XRT} data points systematically higher above 2.5\keV.

Despite the lower energy resolution of \textit{Swift XRT} and the 10 per cent uncertainty in the cross-calibration, inclusion of the \textit{Swift} spectrum is sufficient to constrain the ionisation parameter to $\xi < 4.9$\ergcmps and hence the reflection fraction to $R =  0.34_{-0.12}^{+0.15}$. The photon index is better constrained to $\Gamma = 2.49_{-0.07}^{+0.08}$ and all other parameters are found to be consistent between fitting the \textit{NuSTAR} and \textit{Swift XRT} spectra simultaneously and the \textit{NuSTAR} spectrum alone.

To investigate the cause of the discrepancy between the \textit{Swift} and \textit{NuSTAR} spectra above 2.5\keV, we instead freeze the cross-normalisation constant to 1, hence assuming there to be no discrepancy in the absolute flux calibration between the \textit{Swift} and\textit{NuSTAR} observations and repeat the fit as shown in Fig.~\ref{simfit.fig:noconst}. We then find there to be no discrepancy between the \textit{Swift} and \textit{NuSTAR} observations, and the best-fitting model, between 2.5 and 5\keV. A good fit is still found, albeit marginally worse ($\chi^2 / \nu = 1.04$ with 667 degrees of freedom) and we now find \textit{Swift} data to be slightly below the model (though within the uncertainty of the data points).

We can understand the origin of this discrepancy in terms of the spectral evolution on the decline of the flare. We find that the X-ray flux reflected from the accretion disc is low during the flare and as the flare declines, the reflection fraction increases again to its pre-flare level. The \textit{Swift} observation lasted only 1.4\ks\ and fell early in the 75\ks\ time period spanned by the \textit{NuSTAR} observation, hence will underestimate the reflection fraction compared to the \textit{NuSTAR} observation as a whole. The fit to the \textit{Swift} data is dominated by the reflection-dominated band below 2.5\keV\ rather than energies above 2.5\keV, dominated by the X-ray continuum, where the photon count is lower and there are fewer grouped data points. Hence, when a free cross-normalisation constant is included, this can be used to normalise the weakened soft excess seen in the early \textit{Swift} observation, decreasing the relative model flux in the \textit{Swift} band such that the soft excess measured by \textit{Swift} falls in line with the stronger reflection in the best-fitting model to the \textit{NuSTAR} data. This, however, means that the model now under-predicts the continuum spectrum measured by \textit{Swift}. On the other hand, when the constant is frozen at 1, the X-ray continuum over the energy range 2.5-5\keV\ is in good agreement between the \textit{Swift} and \textit{NuSTAR} spectra as expected from the little variation seen in the spectral slope over the course of the \textit{NuSTAR} observation suggested by the constancy of the hardness ratio shown in Fig.~\ref{nustar_lc.fig}. This interpretation is confirmed by fitting the \textit{Swift XRT} spectrum simultaneously with just the first 5\ks\ of the \textit{NuSTAR} spectrum, selecting a short time interval to be closer to the \textit{Swift} observing time. In this case, the spectra are better matched, with the cross-normalisation constant found to be $1.18_{-0.19}^{+0.26}$, statistically consistent with unity; \textit{i.e.} a statistically significant discrepancy between the \textit{Swift} spectrum and the first 5\ks\ of the \textit{NuSTAR} spectrum between 2.5 and 5\keV\ is not detected. There are insufficient counts, however, in such a short \textit{NuSTAR} exposure to place meaningful constraint on the reflection spectrum.

Fig.~\ref{simfit.fig:ratios} shows the ratio of the observed \textit{Swift XRT} spectra on the day of the flare compared to that the day before and the day after to the best-fitting power law to the \textit{Swift} spectrum above 2\keV\ on each day. It is clear that as the flare declines, the soft excess becomes stronger confirming, in this model, the increase in the reflection fraction in addition to any change in the primary X-ray continuum. The flux in the continuum-dominated band above 2\keV\ falls steadily by around 25 per cent over the three days. Since the reflection fraction is inferred to be increasing as the flare declines, an upper limit to the reflection fraction during the \textit{NuSTAR} observation can be estimated by instead fitting it simultaneously with \textit{Swift XRT} spectrum taken the day after. Including a cross normalisation constant such that the \textit{Swift} and \textit{NuSTAR} observations agree above 3\keV\ (thus increasing the apparent flux in the \textit{Swift} spectrum to account for the decline in flux after the flare), we find the reflection fraction to be $R=0.5\pm 0.2$, consistent with that found fitting the contemporaneous \textit{Swift} spectrum with that from \textit{NuSTAR} and confirming that the reflection fraction is indeed significantly less than unity.

Due to the consistency seen in the band dominated by the X-ray continuum between the \textit{Swift} and \textit{NuSTAR} observations, we adopt the model where the cross-normalisation constant is frozen at 1. The X-ray continuum in this model is found to be slightly harder, with $\Gamma = 2.43_{-0.09}^{+0.05}$ and the 0.1-100\keV\ reflection fraction is found to be $R =  0.41_{-0.15}^{+0.15}$. The best fitting model parameters are shown in Table~\ref{fit_results.tab} compared with those of the same model fit to the 2013 \textit{Suzaku} low state spectrum of Mrk~335 as well as the spectrum from the third of the 2013 \textit{NuSTAR} observations in which the \textit{Swift XRT} count rate was seen to rise from the low state, though for a complete discussion of these earlier spectra, we defer to \citet{gallo+14}, \citet{mrk335_corona_paper} and \citet{parker_mrk335}.

Given that the reflection fraction is inferred to have increased over the course of the \textit{NuSTAR} observation while the flare was subsiding, that the ionisation parameter is low (producing the minimum possible reflected flux below 1\keV) and that the all model parameters are consistent with those found fitting only the \textit{NuSTAR} spectrum, we conclude that the simultaneous fit is sufficient to constrain the upper limit to the average reflection fraction over the course of the observations which, even when considering the \textit{Swift} observation taken the day after the \textit{NuSTAR} observation, is found to be less than unity.

\begin{table*}
\centering
\caption{\label{fit_results.tab}The best fitting model parameters to the \textit{NuSTAR} spectrum of Mrk~335 on the decline of the flare in 2014 fit both independently over the energy range 3-50\keV\ and simultaneously with the 0.3-5\keV\ \textit{Swift XRT} spectrum of Mrk~335. The \textit{NuSTAR} and \textit{Swift} spectra are fit simultaneously with a free cross-normalisation constant between their total photon counts and also freezing this constant to unity to understand the origin of discrepancies arising between 2.5 and 5\keV\ between the two spectra. Considering the time-variability of the \textit{NuSTAR} spectrum, it was determined that the fit in which the constant is frozen to unity (right column) provides the most robust description of the data. The best fitting parameters are compared to those obtained fitting the same model to the 2013 low state spectrum taken with \textit{Suzaku}, taken to be representative of the baseline low state before the flare \citep{mrk335_corona_paper} and the spectrum from the third part of the 2013 \textit{NuSTAR} observation as the count rate increased from the low flux state \citep{parker_mrk335}, refitted using the model employed for the other observations. Also shown is the total X-ray flux calculated from the model over the 0.5-50\keV\ energy band spanned by \textit{Swift XRT} and \textit{NuSTAR} to compare the measured flux between the observations. The corresponding intrinsic luminosity is calculated assuming $\Omega_\mathrm{M}=0.315$ and $\Omega_\Lambda=0.692$, with $H_0 = 67.3$\kmpspMpc \citep{planck_cosmology}. $^f$ denotes parameters frozen during the fit. Reflection fractions, $R = F_\mathrm{ref} / F_\mathrm{cont}$, are shown extrapolated from the model over the energy bands shown, with the flux of the power law continuum (from which $R$ is calculated) and the unblurred reflection also extrapolated over the 0.1-100\keV\ energy range.}
\def\arraystretch{1.5}
\begin{tabular}{llccccc}
  	\hline
   	\textbf{Component} & \textbf{Parameter} &  \textbf{2013 Suzaku} & \textbf{2013 NuSTAR} & \textbf{2014 NuSTAR} & \multicolumn{2}{c}{\textbf{2014 Swift + NuSTAR}}  \\
	\hline
	 \multicolumn{2}{l}{Cross Normalisation NuSTAR/Swift} & -- & -- & -- & $1.3_{-0.2}^{+0.1}$ & $1^f$ \\
	 \hline
	 0.5-50\keV\ Flux & $\lg (F / \mathrm{erg}\,\mathrm{cm}^{-2}\,\mathrm{s}^{-1})$ & $-10.62_{-0.02}^{+0.02}$ & $-10.64_{-0.01}^{+0.05}$ & $-10.59_{-0.03}^{+0.08}$ & $-10.65_{-0.01}^{+0.01}$ & $-10.65_{-0.01}^{+0.01}$ \\
	  & $\lg (L / \mathrm{erg}\,\mathrm{s}^{-1})$ & $43.60_{-0.02}^{+0.02}$ & $43.58_{-0.01}^{+0.05}$ & $43.63_{-0.03}^{+0.08}$ & $43.57_{-0.01}^{+0.01}$ & $43.57_{-0.01}^{+0.01}$ \\
	\hline
	Galactic abs. & $N_\mathrm{H}$ / $10^{20}$\pcmsq & $3.6^f$ & $3.6^f$ & $3.6^f$ & $3.6^f$ & $3.6^f$ \\
	\hline
	Outflow & $N_\mathrm{H}$ / $10^{23}$\pcmsq & $>8.6$ & $2.4_{-1.2}^{+1.7}$ & $1.9_{-0.9}^{+2.1}$ &$1.6_{-0.8}^{+1.6}$ & $1.4_{-0.4}^{+1.4}$  \\
	& $\lg\xi$ / \ergcmps & $3.30_{-0.05}^{+0.05}$ & $3.09_{-0.12}^{+0.15}$ & $2.92_{-0.21}^{+0.17}$ & $2.85_{-0.15}^{+0.17}$ & $2.55_{-0.05}^{+0.08}$  \\
	\hline
	\textsc{Power law} & $\Gamma$ & $1.91_{-0.07}^{+0.04}$ & $2.27_{-0.15}^{+0.10}$ & $2.5_{-0.2}^{+0.2}$ & $2.49_{-0.07}^{+0.08}$ &$2.43_{-0.09}^{+0.05}$  \\
	& $\lg (F / \mathrm{erg}\,\mathrm{cm}^{-2}\,\mathrm{s}^{-1})$ & $-11.82_{-0.13}^{+0.07}$ & $-10.56_{-0.07}^{+0.06}$ & $-10.4_{-0.1}^{+0.2}$ & $-10.48_{-0.09}^{+0.08}$ & $-10.38_{-0.05}^{+0.06}$  \\
	\hline
	\textsc{kdblur2} & $i$ / deg & $58_{-6}^{+4}$ & $57.1^f$ & $57.1^f$ & $57.1^f$ & $57.1^f$   \\
	& $r_\mathrm{in}$ / \rg & $1.25_{-0.02}^{+0.03}$ & $1.67_{-0.09}^{+0.11}$ & $1.235^{+0.9}$ & $1.235^{+0.16}$ & $1.235^{+0.15}$  \\
	\hline
	\textsc{reflionx} & $A_\mathrm{Fe}$ / solar& $6.7_{-1.4}^{+0.8}$ & $2.9_{-1.1}^{+1.0}$ & $2.0_{-0.4}^{+1.4}$  & $5_{-1}^{+3}$ &  $5_{-1}^{+3}$  \\
	& $\xi$ / \ergcmps & $13_{-5}^{+7}$& $<214$ & $<53$ & $< 3.2$ & $< 3.2$  \\
	& $R_{\,0.1-100\,\mathrm{keV}}$ & $14_{-3}^{+6}$ & $<2.5$ & $0.4_{-0.3}^{+0.9}$ & $0.34_{-0.12}^{+0.15}$  & $0.41_{-0.15}^{+0.15}$ \\
	& $R_{\,20-40\,\mathrm{keV}}$ & $36_{-10}^{+12}$ & $3_{-1}^{+3}$ & $1.3_{-0.8}^{+1.9}$ & $4_{-2}^{+4}$ & $4_{-2}^{+4}$  \\
	\hline
	Unblurred refl. & $F$ / $\mathrm{ph}\,\mathrm{cm}^{-2}\,\mathrm{s}^{-1}$ & $1.2_{-0.2}^{+0.2}\times 10^{-3}$ & $7_{-6}^{+4}\times 10^{-4}$ & $<5\times 10^{-4}$ & $3_{-2}^{+2}\times 10^{-4}$ & $3_{-2}^{+2}\times 10^{-4}$  \\
	\hline
	Goodness of fit & $\chi^2 / \nu$ & 1.04 & 0.97 & 1.02 & 1.02 & 1.04  \\
	\hline
\end{tabular}
\end{table*}

\subsection{Accretion disc emissivity profile}

The emissivity profile of the accretion disc (its radial pattern of illumination by the coronal X-ray source) was measured following the method of \citet{1h0707_emis_paper}. The relativistically broadened iron K$\alpha$ emission line was divided into the contributions from reflection at successive radii on the disc (each of which is subject to different gravitational redshift and Doppler shift from a different orbital velocity) and the contribution of each (and hence the reflected flux at each radius) was found through maximum likelihood fitting to the \textit{NuSTAR} spectrum over the energy bands 3-10\keV\ and 3-5\keV.

\begin{figure}
\centering
\includegraphics[width=85mm]{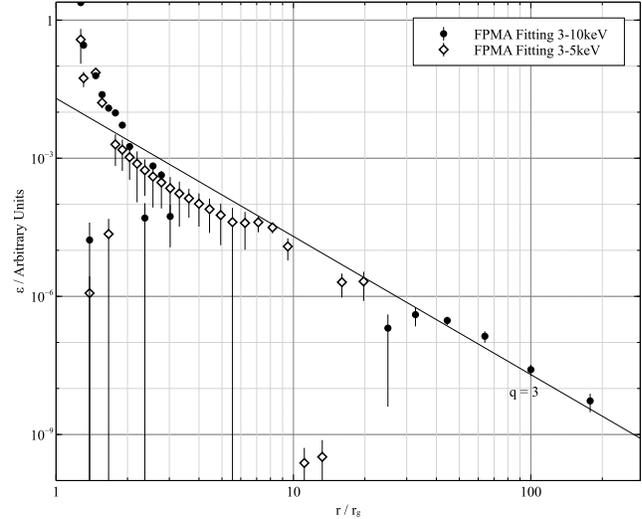}
\caption[]{The emissivity profile of the accretion disc in Mrk~335 on the decline of the 2014 X-ray flare measured from the \textit{NuSTAR} FPMA spectrum. The relativistically blurred reflection from the disc is decomposed into the contributions from successive radii and the best-fitting normalisation of each component to the spectrum is found fitting over both the 3-10 keV and 3-5 keV energy bands. The measured profile is compared to the classical case of a single power law where $\epsilon\propto r^{-3}$.}
\label{emissivity.fig}
\end{figure}

The measured emissivity profile is shown in Fig.~\ref{emissivity.fig}. The accretion disc emissivity falls off steeply as $r^{-6}$ over the inner part of the accretion disc and tending to $r^{-3}$ over the outer disc. \citet{understanding_emis_paper} show that such an emissivity profile is indicative of illumination by a compact corona. There is some evidence for flattening of the emissivity profile over the middle part of the accretion disc which may indicate that the X-ray emitting region is slightly extended over the surface of the disc. Fitting an emission line with a continuous twice-broken power law emissivity profile to the 3-10\keV\ spectrum, we find an upper limit of the source extent to 5\rg\ over the disc at the 90 per cent confidence level.

For comparison with the analysis of \citet{parker_mrk335}, we also fit the 3-50\keV\ \textit{NuSTAR} and 0.3-5\keV\ \textit{Swift XRT} spectra with the \textsc{relxilllp} model which combines the rest-frame reflection spectrum modelled by \textsc{xillver} \citep{garcia+2010,garcia+2011,garcia+2013} with relativistic blurring by the \textsc{relconvlp} model that computes the relativistic blurring from an accretion disc with an emissivity profile appropriate for illumination by an isotropic point source (a `lamppost') at a variable height above the disc plane \citep{dauser+13}. Model parameters are consistent within error to those found using \textsc{reflionx}, with low disc ionisation ($\lg\xi<0.8$) and a best-fitting reflection fraction (which is a free parameter, defined over the 20-40\keV\ energy band in \textsc{relxilllp}) of $6_{-3}^{+2}$, consistent with the value measured using \textsc{reflionx} and \textsc{kdblur2} to model the blurred reflection, shown in Table~\ref{fit_results.tab}. As for the 2013 \textit{NuSTAR} observation, the best-fitting height of the illuminating point source was found to be low, with an upper limit of 2.1\rg\ from the black hole, consistent with our interpretation of a compact source illuminating the accretion disc.

\section{Discussion}
We find that the X-ray spectrum of Mrk~335 observed by \textit{NuSTAR} and \textit{Swift} on the decline of a flare during which the count rate increased by an order of magnitude is well described by the relativistically blurred reflection of a power law X-ray continuum from the surface of the accretion disc, consistent with the configuration seen during the low flux state seen in 2013 by \textit{Suzaku} and \textit{NuSTAR} \citep{gallo+14,mrk335_corona_paper,parker_mrk335}. The X-ray spectrum softened during the flare, as is apparent in the evolution of the hardness ratio measured by \textit{Swift}. This is due to the X-ray continuum from the corona having softened with $\Gamma = 2.43_{-0.09}^{+0.05}$ during the \textit{NuSTAR} observation on the decline of the flare (compared to $\Gamma = 1.91_{-0.07}^{+0.04}$ in the low flux state observed by \textit{Suzaku} in 2013). The X-ray emission becomes much softer during the flare, and softer than it did during the weaker flare in 2013. Such softening of the spectrum as the X-ray emission as the count rate increases has been observed in a number of AGN \citep[\textit{e.g.}][]{taylor_uttley_mchardy}. It can be interpreted as the cooling of the corona with a reduction in energy of each of the scattering particles that produces the X-ray continuum (whether in its own right or due to enhanced cooling by a greater number of scatterings with seed photons) or by a reduction in the optical depth through the corona experienced by the seed photons \citep{1h0707_var_paper}.

The ionisation of the material in the accretion disc remains low, with an upper limit of 3.2\ergcmps, and there is no evidence for truncation of the inner disc, with gravitationally redshifted emission in the relativistically blurred reflection spectrum still detected from $1.235^{+0.16}$\rg, the innermost stable circular orbit for a maximally spinning Kerr black hole. The ionised outflow seen during the 2013 \textit{Suzaku} low state observation \citep{gallo+14} and the 2009 \textit{XMM-Newton} intermediate state observation \citep{longinotti+13} is again detected through narrow absorption features imprinted on the X-ray spectrum in the iron K band.

The similarity of the \textit{NuSTAR} spectrum with that seen as the count rate increased in 2013 \citep{parker_mrk335} is striking with the spectra overlapping almost exactly, save for a slight change in the spectral slope. The softer X-ray emission seen by \textit{Swift}, however, reveals difference between the two observations with approximately five times more photon counts seen in the \textit{XRT} at 0.5\keV\ during the 2014 observations than in 2013. This is accounted for by the softer continuum spectrum in the more recent observations, while the similarity above 5\keV\ is due to the dominance of the reflected spectral component at higher energies, which remains almost constant between the observations (it is the reflection fraction, that is the reflected flux as a fraction of that directly observed in the continuum that drops during the flare). Indeed, fitting the 2014 \textit{NuSTAR} spectrum with the best-fitting model to the 2013 \textit{Suzaku} observation, allowing only the continuum component to vary provides a good fit ($\chi^2 / \nu = 1.02$). Between the best-fitting models, we find that the reflected flux (extrapolated from the observed spectrum over the 0.1-100\keV\ energy band) has decreased by 30 per cent from the 2013 low flux observation to the 2014 flare. Little variation in the reflected flux of the accretion disc has often been seen in AGN as the continuum flux varies greatly and can be explained by either a contraction or movement of the X-ray source closer to the black hole as the continuum count rate decreases causing a greater fraction of the emission to be focused onto the accretion disc \citep{miniutti+04, 1h0707_var_paper}. This constancy of the hard X-ray spectrum above 10\keV\ with variability at lower energies was also seen comparing the 2006 high state and 2013 low state observations of Mrk~335 with \textit{Suzaku}, with the 15-40\keV\ spectrum measured using the PIN hard X-ray detector consistent between the two observations \citep{gallo+14}.

\subsection{The Source of the X-ray Flare}

The measured emissivity profile of the accretion disc is indicative of the accretion disc being illuminated by a compact coronal X-ray source. We find an upper limit of 5\rg\ to the extent of the corona over the surface of the accretion disc, consistent with the findings of \citet{parker_mrk335} and \citet{mrk335_corona_paper} during the 2013 observations (although it is not possible to find a lower limit in this instance since at 4\rg\ the flattening of the emissivity profile due to the extent of the corona is overcome by the steepening of the profile by relativistic effects on the inner disc as discussed by \citealt{understanding_emis_paper}).

Such a compact corona, however, should produce a high reflection fraction with many more photons emitted from the corona focused towards the black hole and hence onto the inner regions of the disc than are able to escape to be observed as part of the continuum \citep{mrk335_corona_paper,1h0707_jan11}. This is discrepant with the low reflection fraction measured (extrapolated from the best-fitting model over the 0.1-100\keV\ energy band) of $R =  0.41_{-0.15}^{+0.15}$. Such a low reflection fraction is not na\"ively expected from a corona illuminating an accretion disc, since even in the classical case, an accretion disc extending some large distance beneath an X-ray source in flat, Euclidean space should receive and reflect half of the rays emitted from the corona, giving $R=1$. While the reflection fraction can be enhanced by the focusing of X-rays onto the inner part of the disc in the strong gravity around the black hole, a diminished reflection fraction can be explained by relativistic motion of the X-ray source upwards causing the emission to be beamed away from the accretion disc \citep{beloborodov}.

In this scenario, we are able to reconcile the measurement of a steep emissivity profile with the low reflection fraction. \citet{mrk335_corona_paper} find that after the weaker flare seen during the 2013 low flux state observation of Mrk~335, the corona became much more compact than before the flare. It is possible that this much greater flare was caused by the vertical collimation of the corona into a `jet-like' configuration and ejection of material as seen in both the 2013 flare and the 2006 high flux state of Mrk~335 observed by \textit{Suzaku} \citep{mrk335_corona_paper}. If, on the decline of the flare, the lower part of the corona had again collapsed down to a confined region around the black hole, this would provide the majority of X-rays that illuminate the accretion disc, giving a steep emissivity profile. Since this flare was much brighter than that in 2006, X-ray continuum emission is still seen from the ejected material, which being both beamed and further from the disc surface is reflected to a much lesser extent, explaining the simultaneous finding of low reflection fraction.

Starting from only the assumption that the observed X-ray spectrum arises from the relativistically blurred reflection from the accretion disc of an X-ray continuum arising from an energetic corona, we find that the best-fitting parameters of such a model indicate the vertical collimation (from the accretion disc emissivity profile) and ejection (from the low reflection fraction) of the corona during the flare. This is combined by a steepening of the X-ray continuum leading to additional flux in the soft X-ray band while the hard X-ray spectrum measured by \textit{NuSTAR} remained constant. The initial assumptions are in line with previous findings that such a model not only reproduces the observed spectra of Mrk~335 through different epochs \citep{crummy+06,larsson+08,gallo+13,parker_mrk335} but naturally explains the changes between high and low flux states \citep{gallo+14,mrk335_corona_paper}.

There could, however, be other possible interpretations of the observed spectra and the cause of the flare. One possibility is that the low reflection fraction is an artefact of additional absorbing material present during the flare (\textit{e.g.} if during the flare an outflow was driven from the inner regions). The additional outflowing material would absorb softer X-rays, reducing the flux below 1\keV, leading to an underestimate of the soft excess above the continuum and, hence, the reflection fraction. For additional absorption of soft X-rays to be accompanied by an overall increase in the soft X-ray count rate measured by \textit{Swift}, the continuum source must intrinsically brighten. The consistency of the \textit{NuSTAR} hard X-ray spectrum with the 2013 observations suggest that this scenario is unlikely without fine-tuning of both the continuum photon index and absorber to keep the hard X-ray spectrum constant. Beginning with the best-fitting model to the 2013 low state observation and freezing the reflection fraction, the 2014 spectrum was fit with additional absorption provided by the \textsc{tbabs} model. We find that allowing the overall model normalisation and the column density of the absorbing material to vary from the low state cannot reproduce the 2014 spectrum with $\chi^2 / \nu = 3.6$. The column density required to reproduce the X-ray flux below 1\keV\ introduces significant curvature into the continuum spectrum and greatly underestimates the observed flux between 2 and 6\keV, hence we conclude that the simple addition of absorption with a brightening of the intrinsic continuum is insufficient to produce the observed spectrum during the 2014 flare.

The flare could also have been caused by an additional source of soft X-ray emission, for instance `soft Comptonisation' by warm, optically thick material associated with the accretion disc \citep{done+12} if such a layer of warm material were to have arisen for a short period but then subsided. We note, however, that since the ionisation parameter is low ($\xi < 3.2$\ergcmps), the model predicts the minimum possible contribution of the reflection spectrum to the soft X-ray band that is admitted by the observed spectrum in the hard, \textit{NuSTAR} band. To confirm this, we fit only the \textit{NuSTAR} spectrum freezing the slope of the continuum spectrum, the photon index $\Gamma$, to the lower statistical bound found previously ($\Gamma=2.25$). This provides the flattest continuum spectrum and will admit the greatest possible contribution from further spectral components in the soft X-ray band. Extrapolating this best-fitting model to the \textit{Swift} band, we find that the measured spectrum is reproduced below 0.7\keV\ and is in-fact slightly overestimated by around $2\sigma$ (before fitting to the \textit{Swift} simultaneously) between 0.7 and 2.0\keV. It is therefore difficult to envisage the flare being caused by an additional emission component in the soft X-ray band, such as Comptonisation by an additional layer of warm material, without too changing the spectral components contributing above 3\keV. Indeed, adding a soft Comptonisation component using the \textsc{optxagnf} model \citep{done+12} in addition to the power law continuum and relativistically blurred reflection required to fit the iron K$\alpha$ line and Compton hump seen in the \textit{NuSTAR} spectrum, we find that fit is not improved with the extra spectral component ($\Delta\chi^2 = 1$ but with 4 fewer degrees of freedom). The low reflection fraction ($R=0.42_{-0.17}^{+0.13}$) and softer continuum ($\Gamma = 2.40_{-0.11}^{+0.06}$) are still required and at its upper statistical limit (at the 90 per cent confidence level), the soft Comptonisation component contributes less than half of the X-ray flux of the blurred reflection to the X-ray spectrum. To allow for the most general case and to find the broadest statistical limits, the \textsc{optxagnf} model was fit with the radius and optical depth of the corona as free parameters and the temperature of the corona producing the soft Comptonised component free, but limited to a maximum of 0.5\keV\ such that it only contributed to the soft excess above the power law seen in the \textit{Swift} spectra. Thus, we conclude that a change in geometry and motion of the corona that produces the primary X-ray continuum and illuminates the disc is the most likely explanation for the flare.

It should be noted that while the soft X-ray count rate rises sharply during the flare, the total X-ray flux over the broad 0.5-50\keV\ energy band remains roughly constant, within uncertainty. This suggests that while the geometry of the X-ray emitting corona is changing in this interpretation, with the collimation and ejection of its constituent material, the total energy output of the corona through X-rays remains constant. The flare seen in the \textit{Swift XRT} lightcurve is due to the softening of the continuum spectrum with more continuum photons being produced but with lower average energy, coupled with the drop in reflection fraction due to the change in geometry and relativistic beaming.

\subsection{The UV Variability}
Variability in the emission from accreting black holes is thought to be based on fluctuations of the mass accretion rate through the disc with the accretion of over-dense regions of the disc increasing the gravitational binding energy that is liberated as radiation. Such an overdensity in the accretion disc propagates inwards through a viscously heated accretion disc, such as the standard $\alpha$-disc model of \citet{shaksun}. Since energy is liberated locally in the disc by viscous heating, elements of material at each radius in the disc radiate as a black body and the accretion of the overdensity leads to an increase in the thermal emission at progressively shorter wavelengths as it reaches the hotter, inner parts of the disc. The thermal emission from accretion discs around supermassive black holes in AGN peak at ultraviolet (UV) wavelengths. It is not until the overdensity reaches the innermost regions where the accretion flow is coupled to the X-ray emitting corona that the corresponding increase would be seen in the X-ray continuum emission. In order to understand the variability seen in the UV emission, shown in Fig.~\ref{swift_lightcurve.fig}, we hence consider the propagation of an overdensity through a viscously heated disc.

We can estimate the time-scale on which such an overdensity would propagate inwards. In a standard Shakura-Sunyaev accretion disc, the local viscous velocity at which material travels radially inward is
\begin{equation}
\dot{r} = r^{-\frac{1}{2}} \frac{\alpha}{\sqrt{GM}}\left(\frac{h}{r}\right)^2
\end{equation}
Which can be integrated to give the travel time of the overdensity between two radii. The values of the $\alpha$ viscosity parameter and the disc aspect ratio $(h/r)$ are unknown, however are often assumed to follow a power law with radius, $\alpha(h/r)^2 = Cr^{-\beta}$ with the constant C set to reproduce a thick disc with $\alpha = 0.3$ and $(h/r)=1$ on the inner edge \citep{arevalo+2006}. Given that an increased value of $\beta$ leads to slower propagation through the outer disc, we can estimate an upper limit by noting that the X-ray flare declines sharply, within 8 days. Given the corona is measured to be compact, we infer that the overdensity must cross the innermost 10\rg, where we assume the disc to be coupled to the corona, within this time. This time-scale is plausible in this model and requires $\beta < 2$ for a black hole mass $2.6\times10^7$\Msun\ as measured for Mrk~335 \citep{grier+12}.

In the classical approximation, the temperature profile of a Shakura-Sunyaev accretion disc due to black body emission from local dissipation by viscous heating of the accreting material is given by
\begin{equation}
\label{disctemp.equ}
T(r) = \left[ \frac{3GM\dot{M}}{8\pi\sigma_\mathrm{SB} r^3} \left(1 - \sqrt{\frac{r_\mathrm{in}}{r}} \right) \right]^\frac{1}{4}
\end{equation}

Writing the accretion rate, $\dot{M}$, as a fraction of that required to achieve the Eddington luminosity $\dot{M}_\mathrm{Ed}$ and assuming a rest mass to luminosity conversion efficiency $L = 0.1\dot{M}c^2$, we can write
\begin{align*}
T(r) = &\left( 1.1\times 10^6 \right) \left( \frac{M}{10^8 M_\odot} \right)^{-\frac{1}{4}} \left( \frac{\dot{M}}{\dot{M}_\mathrm{Ed}} \right)^{\frac{1}{4}} \\
& \left( \frac{r}{r_\mathrm{g}} \right)^{-\frac{3}{4}} \left( 1 - \sqrt{\frac{r_\mathrm{in}}{r}}\right)^\frac{1}{4}\,\mathrm{K}
\end{align*}

The black body spectrum radiated from each radius peaks at a frequency $h\nu \approx 2.8k_\mathrm{B}{T}$, thus for a black hole of mass $2.6\times10^7$\Msun\ accreting at 0.2 times the Eddington rate (to maintain a geometrically thin accretion disc from which reflection can be observed), the UV emission in the \textit{Swift UVW2} bandpass will be dominated by thermal emission from around 100\rg\ in the disc. The upper limit (setting $\beta=2$) on the propagation time of an overdensity from this radius at which UV emission in the \textit{UVW2} bandpass peaks to within $\sim 10$\rg, where the disc is likely to be coupled to the corona, is around 20 years. We can hence rule out that the variability in UV emission seen from the dip around 80 days prior to the X-ray flare to the peak (around MJD\,56825), coinciding and then subsiding with the flare is the excess thermal emission from an overdensity in the disc that causes the increased coronal X-ray emisison during the flare. The thermal emission from the accretion disc within the X-ray emitting region peaks at around 150\,\AA, hence variability in the \textit{UVW2} bandpass cannot be connected to contemporaneous variability in the X-ray emission simply through thermal emission from corresponding overdensities in the accretion disc.

It does, however, appear that there is a modest increase in the UV emission connected with the X-ray flare with the peak UV flux coinciding with the peak X-ray count rate and the two subsiding together. The UV flux measured through the \textit{UVW2} filter increases by around 25 per cent, compared to a 10 times increase in the X-ray count rate. This is unlikely to be due to the thermal reprocessing of the excess continuum emission seen during the flare from the accretion disc since the X-ray flux reflected from the accretion disc remains relatively constant between the 2013 observations and the 2014 flare, suggesting there is not a significant increase in the continuum flux irradiating the disc.

The observed rise in UV emission during the X-ray flare could, however, be emission from the jet-like outflowing corona that is inferred from both the steep emissivity profile of the accretion disc and the low reflection fraction suggesting beaming of the continuum emission away from the accretion disc. While the X-ray continuum is produced by the Compton up-scattering of thermal seed photons from the accretion disc by energetic particles within the corona, it is plausible that emission is generated within this outflowing material. Thermal emission from this material is unlikely to be detected in the \textit{UVW2} bandpass. Material ejected from the inner regions of the accretion flow will likely be at least as hot as the inner regions of the disc, which from Equation~\ref{disctemp.equ} would produce a black body spectrum peaking around 150\,\AA. The black body flux would be lower by a factor of 1000 in the \textit{UVW2} bandpass.

A corona outflowing at a relativistic velocity through a magnetised region is expected to produce synchrotron radiation peaking at radio wavelengths which can also be up-scattered by the energetic particles. This synchrotron self-Compton emission has long been invoked to explain the emission seen from jets in blazars \citep[\textit{e.g.}][]{ghisellini+85} and can readily produce UV emission in the \textit{UVW2} band for even the modest bulk Lorentz factors ($\Gamma\sim 1$, corresponding to $v\sim 0.3c$) expected to produce the measured reflection fraction of 0.3 \citep{beloborodov,gallo+14}. We expect the detected emission to be weak compared to the thermal emission from the outer parts of the accretion disc since the measured 57\,deg inclination of the accretion disc suggests we are not looking down what would be the jet axis, although the opening angle of beamed synchrotron emission ($\Delta\theta\sim\gamma^{-1}$) is relatively large at 50\,deg from an outflow at $0.3c$. This would offer an explanation as to why the 10 times increase in X-ray count rate is accompanied by an increase in \textit{UVW2} flux by only the 25 per cent that is observed.

While the long propagation time-scales mean we cannot rule out the flare being caused by the accretion of an overdense region of the accretion disc, there is the possibility the cause of the flare is manifested on the inner parts of the accretion flow or within the corona that produces the continuum emission. General relativistic magnetohydrodynamic simulations of magnetised accretion flows onto black holes conducted by \citet{mckinney+2012} suggest that large-scale toroidal magnetic fields can, in particular circumstances, be accreted to the inner regions of the accretion disc where the magnetic flux density is greatly increased. The accretion flow becomes magnetically choked as the high flux density on the inner disc compresses the disc and limits the accretion of material, reducing the mass accretion rate. When a critical magnetic flux density is reached, the magnetic field inverts, ejecting material perpendicular to the disc plane and relieving the magnetic flux density choking the accretion flow. If the low flux state Mrk~335 has been observed in over recent years is due to magnetic choking of the accretion flow, \citet{mrk335_corona_paper} suggest that the changes seen in the X-ray emitting corona and the ionisation of the disc during and immediately after the X-ray flare during the 2013 \textit{Suzaku} observation could be due to such a field inversion launching the flare and removing the magnetic field that compresses the inner disc. This much greater X-ray flare might well have been caused by the same phenomenon; the corona becomes vertically collimated and energetic particles from the disc surface are launched by a magnetic field inversion on the inner regions of the accretion disc, giving rise to the excess continuum emission.

\subsection{Putting together the variability of Mrk~335}
\citet{mrk335_corona_paper} show that the variability of Mrk~335 between high and low flux states over the last decade can be understood in the context of the changing geometry of the corona that produces the observed X-ray continuum. In the high flux state, observed by \textit{XMM-Newton} in 2006, the corona was found to be extended over the surface of the accretion disc, while in the 2009 intermediate and 2013 low flux states, the corona was confined to a smaller region around the black hole. Not only had the total X-ray luminosity reduced, but the proximity of the X-ray continuum emission to the black hole meant that more of the emitted rays were focused towards the black hole. This means that more radiation is focused onto the accretion disc rather than being able to escape to be observed as part of the continuum, greatly increasing the reflection fraction. Likewise, the high flux state observed by \textit{Suzaku} in 2006 could be explained by X-ray emission from a more extended corona, except in this case the corona appeared to be extended vertically in a `jet-like' structure while outflowing at a mildly relativistic velocity with relativistic beaming of emission away from the accretion disc explaining the incredibly low reflection fraction (less than unity) in this case. In general, the X-ray continuum emission is seen to be softer, with steeper photon indices in the high flux states compared to the low.

During the 2013 low flux state, a short flare was observed in the X-ray emission with the count rate doubling for around 100\ks. \citet{mrk335_corona_paper} show that this flare can be understood in terms of the collimation of the corona into a slightly vertically extended structure as the count rate increased, causing a drop in the reflection fraction and accompanied by a softening of the continuum spectrum. As the flare dissipated, the corona collapsed into an incredibly confined region within just 2\rg\ of the black hole. The 2014 flare is consistent with having arisen from the same process but on a larger scale. In 2014, the count rate increased by a factor of 10 and the continuum became much softer (hence the greater X-ray count rate by a factor of 4 observed with \textit{Swift} in 2014 compared to 2013). The X-ray spectra on the decline of the flare show a low reflection fraction (lower than in 2013), consistent with the relativistic beaming of emission away from the disc from a faster and more extended outflow. The emissivity profile shows that the disc is illuminated by a compact corona close to the black hole. This is consistent with the vertical collimation and ejection of material in the corona, as for the 2013 flare, but with the greater magnitude of the flare, we are still able to observe the X-ray continuum emitted from the ejected portion of the corona while the base that illuminates the disc has again collapsed to a confined region. In the case of the 2013 and 2014 flares, the jet-like outflowing corona was short-lived, however in the 2006 high state observed by \textit{Suzaku}, this same geometry was sustained for at least the period of the observation.

There is perhaps evidence emerging that long-lived high flux states such as that seen until the 2006 \textit{XMM-Newton} observation in Mrk~335, as well as in the 2002-2010 observations of the NLS1 galaxy 1H\,0707$-$495, correspond to coronae extended over the surface of the accretion disc that have a greater cross-section to up-scatter the disc thermal emission, while short-lived X-ray flares are linked to a vertical collimation (and mildly relativistic outflowing) of the corona. It should be noted that these conclusions are derived from the geometry of the corona inferred from the measured accretion disc emissivity profile and reflection fraction rather than from the application of a specific model of the formation and structure of a corona arising from the accretion flow, although the length of the flare is longer than the viscous time-scale of material passing through the inner 10\rg\ of the accretion disc and much longer than the light crossing time of this region, so the inferred reconfiguration of the corona is plausible in terms of the flare time-scale.

Not only is variability seen in the corona of Mrk~335, but also in the outflow, attributed to a wind from the accretion disc, that is seen to imprint absorption features on the observed X-ray spectrum. The outflows are most prominent in the intermediate flux state observed by \textit{XMM-Newton} in 2009 \citep{longinotti+13} but can also be detected in the 2013 low flux state \citep{gallo+14} and the 2006 (\textit{XMM-Newton}) high flux state in which the corona was extended over the surface of the disc \citep{mrk335_corona_paper}. On the other hand, no such absorption features are observed in the 2006 \textit{Suzaku} observation when the corona had become vertically collimated \citep{larsson+08}. We find that the outflow had weakened slightly, with a factor 4 less column density and lower ionisation during the 2014 observations on the decline of the flare compared to the low flux state observed with \textit{Suzaku} in 2013. It was not, however, possible to detect a significant difference, given the errors, between the absorber during the 2014 flare and during the 2013 \textit{NuSTAR} observation as the flux rose following the \textit{Suzaku} low state observation. The non-detection of the outflow during the 2006 \textit{Suzaku} observation and its weakening during the 2014 flare compared to the previous low flux state could suggest that the apparent dichotomy between the launching of winds from the accretion disc and the launching of jets in X-ray binaries (or in this case, the outflowing and collimation of the corona in a `jet-like' structure) reported by \citet{king_wind_jet} could extend to supermassive black holes with Mrk~335 undergoing such a transition.

\section{Conclusions}
\textit{Swift} and \textit{NuSTAR} spectra were analysed from the decline of an X-ray flare in Mrk~335 lasting around 30 days. At the peak of the flare, the X-ray count rate had increased by a factor of 10 from the previously observed low flux state.

The X-ray emission is largely consistent with the low flux state seen by \textit{Suzaku} in 2013. The most notable changes between this low flux state and the flare are an increase in the directly observed continuum flux and the softening of the continuum spectrum from a photon index of 1.9 to 2.5. The measured emissivity profile of the accretion disc suggests that it is illuminated by a compact X-ray source extending no more than 5\rg\ over the surface of the disc, while a very low reflection fraction is measured with ratio of reflected to directly observed continuum flux of just 0.41.

Combining these observations, we interpret the X-ray flare as arising from the vertical collimation and ejection of the X-ray emitting corona at a mildly relativistic velocity, causing the continuum emission to be relativistically beamed away from the disc. As the flare subsides, the base of this jet-like structure collapses into a compact X-ray source that provides the majority of the radiation that illuminates the disc while continuum emission is still detected from energetic particles further out, maintaining the low reflection fraction.

While the soft X-ray count rate is seen to rise sharply during the flare, the total X-ray flux was found to remain constant between the flare and previous low flux state observations, suggesting that the total X-ray energy output of the corona does not change and it produces more photons with lower average energy during the flare.

The origin of the increase in UV flux coincident with the X-ray flare is explored and it is plausible that this originates from the ejected coronal material through synchrotron self-Compton emission.

\section*{Acknowledgements}
DRW is supported by a CITA National Fellowship. This work is based on observations made by the \textit{NuSTAR} mission, a project led by the California Institute of Technology, managed by the Jet Propulsion Laboratory, and funded by NASA.

\bibliographystyle{mnras}
\bibliography{agn}

\label{lastpage}

\end{document}